\documentclass[a4paper]{article}

\usepackage[top=2cm, bottom=2cm, left=2cm, right=2cm]{geometry}

\usepackage[dvipdfmx]{graphicx}

\usepackage[colorinlistoftodos]{todonotes}
\usepackage{authblk}
\usepackage{hyperref}
\usepackage{balance}
\usepackage{colortbl}
\usepackage{tabularx}

\title{See-Through Captions: Real-Time Captioning on Transparent Display for Deaf and Hard-of-Hearing People}

\makeatletter
\newcommand{\printfnsymbol}[1]{%
  \textsuperscript{\@fnsymbol{#1}}%
}
\makeatother

\author[1]{Kenta Yamamoto\thanks{Three authors contributed equally to this research.}}
\author[1]{Ippei Suzuki\printfnsymbol{1}}
\author[1]{Akihisa Shitara\printfnsymbol{1}}
\author[1]{Yoichi Ochiai}
\affil[1]{University of Tsukuba}

\date{}

\begin{document}

\maketitle

\begin{figure}[h]
    \centering
    \includegraphics[width=\textwidth]{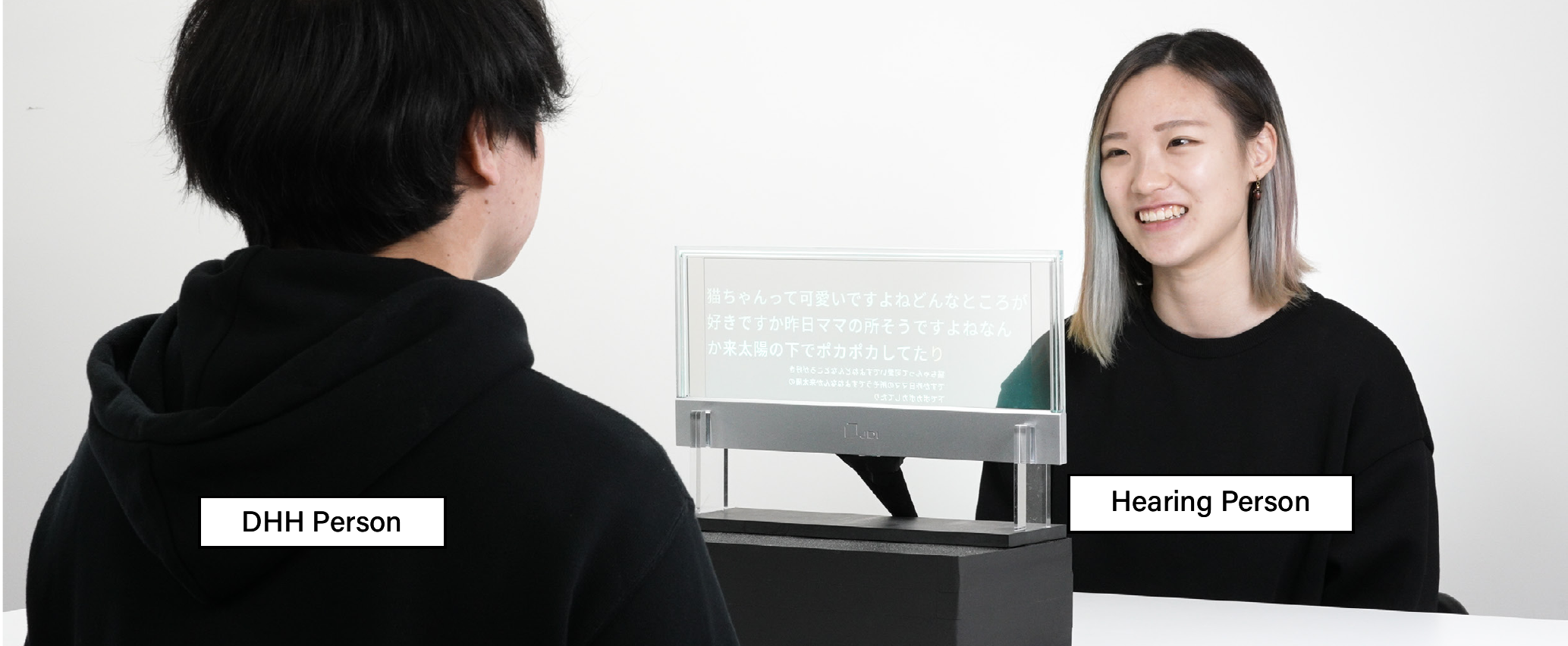}
    \caption{Proposed real-time captioning system.}
    \label{fig:teaser}
\end{figure}

\begin{abstract}
Real-time captioning is a useful technique for deaf and hard-of-hearing (DHH) people to talk to hearing people.
With the improvement in device performance and the accuracy of automatic speech recognition (ASR), real-time captioning is becoming an important tool for helping DHH people in their daily lives.
To realize higher-quality communication and overcome the limitations of mobile and augmented-reality devices, real-time captioning that can be used comfortably while maintaining nonverbal communication and preventing incorrect recognition is required.
Therefore, we propose a real-time captioning system that uses a transparent display.
In this system, the captions are presented on both sides of the display to address the problem of incorrect ASR, and the highly transparent display makes it possible to see both the body language and the captions.
\end{abstract}

\section{Introduction}
Deaf and hard of hearing (DHH) people need to be more accessible to sound information in their daily lives.
The importance of spoken language in interpersonal communication, education, and business situations has necessitated its visualization.
Visualization methods include real-time translation by an interpreter using sign language~\cite{10.1093/deafed/enj007} and real-time spoken-language transcription by a supporter~\cite{10.1145/2380116.2380122}; however, they are expensive and thus infeasible for casual daily use.
In recent years, real-time captioning via automatic speech recognition (ASR) has attracted attention because of the advances in the processing performance of mobile devices, internet-communication speed, and speech-recognition accuracy.
This technology enables the provision of support to DHH people in various situations.

ASR has long been expected to serve as a universal method of accessing voice information~\cite{wald2008universal}.
Significant effort has been devoted to introduce such a technology in the field of education~\cite{10.1145/638249.638284} and implement the use of this technology by DHH people~\cite{10.1145/1268784.1268860, 10.1145/2543578, 10.1145/2982142.2982164}.
The main method used in the educational setting is the capturing of teachers' voice information via ASR.
After studies in a static context such as lectures in a classroom, research on the free utilization of speech recognition by DHH people in various situations has been conducted.
In methods adopted in previous studies, the sound information to be recognized by DHH people was sent to translators via the internet, and the text that they verbalized was sent to mobile devices~\cite{10.1007/11853565_10, 10.1145/1878803.1878885}.
With the development of high-performance mobile devices, such as smartphones and smartwatches, research on text conversion via ASR on mobile devices has increased~\cite{10.1145/3234695.3236362}.
Furthermore, with the progress made in the augmented reality (AR) technology, research on the use of AR to realize caption displays is being actively conducted~\cite{10.1145/3234695.3236362, 10.1145/3197391.3205404, 10.1145/3173574.3173867, 10.1145/3379337.3415817}.

The method of realizing real-time captioning via ASR using such mobile and AR devices has some limitations.
For example, when using a mobile device, the body language of the partner cannot be confirmed because the mobile device must be viewed to see the ASR result; when using an AR device, DHH people can see the voice-recognized text while observing the partner's body language.
Previous studies have favored this advantage of using an AR device~\cite{10.1145/3290605.3300276}.
However, when an AR device is used, the speaker who does not wear it cannot confirm whether the speech has been correctly recognized, which may lead to errors in communication.

In this study, to address these problems, we developed a real-time captioning system that utilizes a transparent display and allows people to check the speech-recognition results while seeing their partner during the conversation (Fig.~\ref{fig:teaser}).
Table~\ref{tb:comparison} summarizes the characteristics of the proposed system and existing methods (mobile devices and AR devices).
Although the proposed system is limited to one-on-one communication, we expect it to help in improving the quality of communication because it can prevent incorrect ASR and overlooking of the body language of the partner.
In addition, the installation of this system in common places where voice conversation is expected, such as cash registers in supermarkets and reception desks at government offices, can possibly help DHH people in their daily lives.

\renewcommand{\arraystretch}{1.25}
\begin{table}[t]
    \small
    \centering
    \caption{Comparison with previous methods of real-time captioning.}
    \label{tab:freq}
    \begin{tabular}{lccc}
        \hline
            & 
            \begin{tabular}{c}
                \textbf{Mobile Device}\\
                (Smartphone, Smartwatch)
            \end{tabular}&
            \textbf{AR Device} &
            \textbf{Transparent Display}
        \\
        \hline
            \begin{tabular}{l}
                Confirmation of ASR result\\
                by hearing person 
            \end{tabular} &
            Yes &
            No & 
            Yes
        \\
        \hline
            \begin{tabular}{l}
                Seeing body language of\\
                hearing person
            \end{tabular} &
            No & 
            Yes & 
            Yes
        \\
        \hline
            \begin{tabular}{l}
                Situation of communication 
            \end{tabular} &
            \begin{tabular}{c}
                Multiple People\\ 
                One-to-One 
            \end{tabular} &
            \begin{tabular}{c}
                Multiple People\\ 
                One-to-One 
            \end{tabular} &
            \begin{tabular}{c}
                One-to-One
            \end{tabular}
        \\
        \hline
    \end{tabular}
    \label{tb:comparison}
\end{table}
\renewcommand{\arraystretch}{1.0}

\section{Proposed System}

\subsection{System Configuration}
The proposed system presents the result of real-time captioning via ASR on a transparent display, and hearing and DHH people can communicate while checking the speech-recognition result.
The two main functions of this system are ASR and caption display.
First, a directional microphone is used as a speech-input device for ASR, and only the speech of the hearing person is used as the input speech.
The input speech is converted into text via ASR.
The system uses SpeechRecognition of the Web Speech API from Google Chrome (ver 87.0.4280.88) for ASR.
Next, the text of the speech-recognition result is presented on the transparent display in real time.
In this prototype, a transparent display from Japan Display Inc.\footnote{\url{https://www.j-display.com/technology/jdinew/clear_display.html} [Last Accessed: 01/11/2021]} is used, and the entire processing from ASR to caption display is performed on the browser.

\subsection{Communication Process}
\begin{figure}[t]
    \centering
    \includegraphics[width=\linewidth]{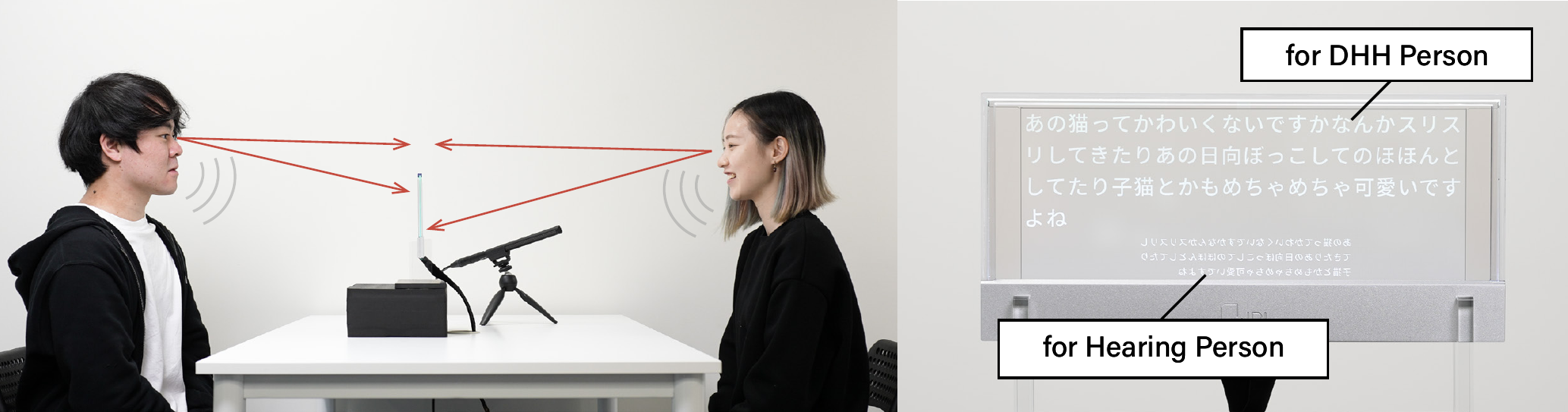}
    \caption{(left) Photograph of DHH person and hearing person using the proposed system. They talk using their voice while seeing the captions displayed on the transparent display. (right) Captions presented on the display. A small caption with left and right flips is displayed for the hearing person.}
    \label{fig:proposed_system}
\end{figure}

The proposed system is expected to be used in one-to-one and face-to-face communication.
The conversation between two people across a transparent display is illustrated in Fig.~\ref{fig:proposed_system}.
A hearing person speaks into a speech-input device, and the recognized speech is converted into text and presented on a transparent display.
The captions are primarily meant for the DHH person to read.
However, to enable the hearing person to confirm that there is no significant difference between their speech and the ASR result, the same text as the caption for the DHH person is displayed with the left and right reversed.
If the ASR result is incorrect, the hearing person indicates through gestures that there was a recognition error and then repeats their speech such that it is correctly recognized.
The DHH person understands the spoken-language content by reading the captions on the transparent display and responding through their voice.
The hearing person listens to the voice of the DHH person and inputs the voice to the microphone again.
The conversation continues with a repetition of these exchanges.
The characteristics of the transparent display allow for the displaying of captions on both sides of a single display.

\subsection{Caption Design}
The caption design has a significant effect on readability.
Previous studies have investigated the caption-appearance preferences~\cite{10.1145/3290607.3312921} and designs based on the reliability of speech recognition~\cite{10.1145/3132525.3132541}.
The proposed system provides an interface that allows the user to change the caption-design parameters such as character size, character color, character transparency, and font.
There are some design parameters that have not been examined; these include the character-display speed, number of lines, and the caption-disappearance speed.
Therefore, further improvements are required in the future.

\section{Design Space of See-Through Caption}

\subsection{How to Design Optimal Real-Time Captioning}
\begin{figure}[t]
    \centering
    \includegraphics[width=\linewidth]{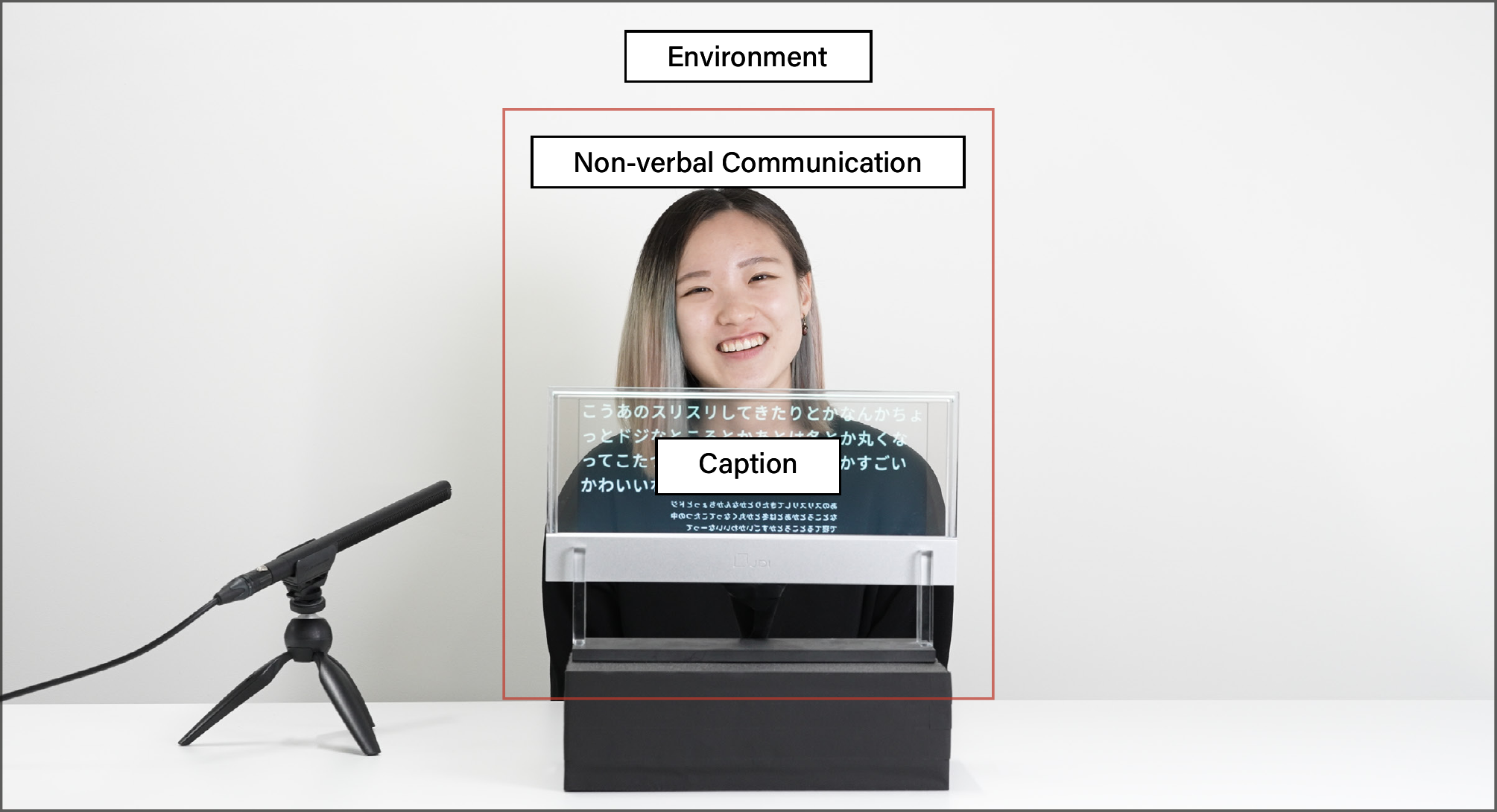}
    \caption{Three factors for superior real-time captioning system with transparent displays. This picture was taken from the viewpoint of a DHH person when using this system.}
    \label{fig:design_space}
\end{figure}

What is the optimal real-time captioning?
We consider three factors to be important for superior real-time captioning, as shown in Fig.~\ref{fig:design_space}.

\subsubsection{Assisting in understanding of spoken language}
Enabling accurate and smooth access of audio information by DHH people is the most important role of real-time captioning.
Therefore, it is necessary to improve not only the speech-recognition accuracy but also the overall quality of captioning, and we must thus consider the design of characters that make up captions and how they can be displayed.

\subsubsection{Not interfering nonverbal communication}
Communication between people is not limited to spoken language.
A previous study emphasized the importance of eye contact during a conversation~\cite{10.1145/3290605.3300276}.
Captioning tools should be designed for nonverbal communication as well.
Through the use of a transparent display, the proposed system avoids any obstruction to nonverbal communication, similar to captioning by AR devices.
In the future, it will be necessary to further consider how captions can be positioned relative to the speaker and how they can displayed such that they do not obstruct nonverbal communication.

\subsubsection{Not spoiling interior design}
Finally, we should consider how the device for real-time captioning contributes to the designing of an indoor space.
As mentioned earlier, the proposed system is expected to be used in daily life (e.g., at cash registers and receptions).
However, in such situations, captioning is only used when necessary.
Therefore, it is also important to integrate the display at the installation location.
Because mobile and AR devices are carried by a DHH person, the location design need not be considered.
Owing to the transparency of the display, visual impediments are less than those in the case of a normal display; however, a detailed discussion will be required.

\subsection{Future Work}
We have developed and demonstrated a see-through type real-time captioning system that utilizes a transparent display.
Shitara, who is a deaf person and one of the authors of this manuscript, was involved in the development and designing of the system.
However, more user studies are required to achieve a better design.
For example, it is necessary to further clarify the merits and demerits of this system from the user studies and carefully compare them with those of mobile and AR devices.
Furthermore, it is necessary to investigate the relationship between the understanding of verbal and nonverbal communication in real-time captioning by performing eye-gaze analysis during conversations.
We believe that this system is useful for supporting the communication with DHH people, and we will conduct appropriate user studies to achieve a better system.

\balance

\bibliographystyle{unsrt}
\bibliography{reference}

\end{document}